\documentclass{article}%
\usepackage[margin=1.55in]{geometry}
\usepackage{amsmath}
\usepackage[linkcolor=blue,urlcolor=blue,colorlinks=true]{hyperref}
\usepackage{amsfonts}
\usepackage{amssymb}
\usepackage{graphicx}
\usepackage{diagbox}
\usepackage{subfig}
\usepackage{epsfig}%
\usepackage{color}
\usepackage[table]{xcolor}
\usepackage{multirow}
\usepackage{setspace}
\usepackage{algorithm}
\usepackage{algpseudocode}
\usepackage{hyperref}
\usepackage{enumitem}
\usepackage{hhline}
\usepackage{pbox}
\usepackage{booktabs}
\usepackage{lscape}
\usepackage{array}
\usepackage[flushleft]{threeparttable}
\usepackage{authblk}
\usepackage{float}
\usepackage{supertabular}
\usepackage[title]{appendix}


\begin{document}
\doublespacing

\title{\textbf{A Bayesian Nonparametric Estimation of Mutual Information}}   

\author[1]{Luai Al-Labadi\thanks{{\em Corresponding author:} luai.allabadi@utoronto.ca}}

\author[2]{Forough Fazeli-Asl\thanks{forough.fazeli@math.iut.ac.ir}}

\author[2]{Zahra Saberi\thanks{ z\_saberi@cc.iut.ac.ir}}

\affil[1]{Department of Mathematical and Computational Sciences, University of Toronto Mississauga, Mississauga, Ontario L5L 1C6, Canada.}
\affil[2]{Department of Mathematical Sciences, Isfahan University of Technology, Isfahan 84156-83111, Iran.}

\date{}
\maketitle

\pagestyle {myheadings} \markboth {} {A  BNP estimation of MI via DP.}

\begin{abstract}
Mutual information is a widely-used information theoretic measure to quantify the amount of association between variables. It is used extensively in many applications such as image registration, diagnosis of failures in electrical machines, pattern recognition, data mining and  tests of independence.
The main goal of this paper is to provide an efficient estimator  of the  mutual information based on the approach of Al Labadi et. al. (2021). The estimator is explored through various examples and is compared to its frequentist counterpart due to  Berrett et al. (2019). The results show the good performance of the procedure by having a smaller mean squared error.

\par

 \vspace{9pt} \noindent\textsc{Keywords:} Dirichlet process, $k$-nearest neighbor distance, Multivariate entropy, Simulation.

 \vspace{9pt}

\end{abstract}
\section{Introduction}
Mutual Information is a measure to quantify the level of dependency between random variables. Specifically, let $\mathbf{X}=(X_{1},\cdots,X_{d})$ be a random vector with joint continuous distribution function $F$ and marginal continuous distribution functions $F_{1},\cdots,F_{d}$. Then mutual information between $X_{1},\cdots,X_{d}$ is defined as
\begin{equation}\label{MI-basic}
MI(F)=\int_{-\infty}^{\infty}\cdots\int_{-\infty}^{\infty} f(x_{1},\ldots,x_{d})\log\dfrac{f(x_{1},\ldots,x_{d})}{f(x_{1})\ldots f(x_{d})}\, dx_{1}\cdots dx_{d},
\end{equation}
where $f(x_{1},\cdots,x_{d})$ and $f(x_{i})$ denote, respectively, the probability density functions of $F$ and $F_{i}$, $i=1,\ldots,d$. Note that, throughout this paper, $\log(\cdot)$ denotes the natural logarithm. Clearly,  \eqref{MI-basic} is the Kullback-Leibler distance of $F$ from the product of $F_i$'s and so it is non-negative. After simplification, \eqref{MI-basic} can be written as
\begin{align}\label{MI-Entropy}
MI(F)&=\int_{-\infty}^{\infty}\cdots\int_{-\infty}^{\infty}f(x_{1},\ldots,x_{d})\log f(x_{1},\ldots,x_{d}) \, dx_{1}\cdots dx_{d}\nonumber\\
&-\sum_{i=1}^{d}\int_{-\infty}^{\infty}f(x_{i})\log f(x_{i})\, dx_{i}\nonumber\\
&=-H(F)+\sum_{i=1}^{d}H(F_{i}),
\end{align}
where $H(F)$ and $H(F_{i})$ denote, respectively,  the entropy of $F$ and $F_{i}$. From the above definition, it  is clear that to estimate $MI(F)$, we need to develop estimators for $H(F)$ and $H(F_i)$. Although many methods of entropy estimation can be found in literature, the works are often related to the univariate (marginal) entropy estimation. For a comprehensive study, see, Al-Labadi et al. (2021) and the references therein. For the multivariate (joint) entropy estimation, some frequentist procedures have been offered in the literature; see, for instance,  Kozachenko and Leonenko (1987), Misra et al. (2010), Sricharan and
Hero (2012), Sricharan et al. (2013), Gao et al. (2016),
Berrett et al. (2019), Ba and Lo (2019) and the references therein. It should be note that,  Kozachenko and Leonenko (1987) (KL) estimator is the most well-known one. It is based on the first nearest neighbor distances of the sample points. A $k$-nearest neighbor ($k$-NN) version of KL estimator is proposed by Singh et al. (2003) to improve the performance of the  estimator.  Let $\mathbf{X}_{1},\cdots,\mathbf{X}_{n}$ be $n$ independent random vectors each having the continuous $d$-variate cdf $F$ and let, for $i=1,\ldots,n$, $R_{i,k,n-1}=||\mathbf{X}_{(k),i}-\mathbf{X}_{i}||$ and $\mathbf{X}_{(1),i},\ldots,\mathbf{X}_{(k),i},\ldots,$ $\mathbf{X}_{(n-1),i}$ is a reordering of $\lbrace \mathbf{X}_{1},\ldots,\mathbf{X}_{n}\rbrace\setminus\lbrace\mathbf{X}_{i}\rbrace$ such that $||\mathbf{X}_{(1),i}-\mathbf{X}_{i}||\leq\ldots\leq||\mathbf{X}_{(k),i}-\mathbf{X}_{i}||\leq\ldots\leq||\mathbf{X}_{(n-1),i}-\mathbf{X}_{i}||$, where $||\cdot||$ denotes the Euclidean norm on $\mathbb{R}^{d}$ and $A\setminus B$ denotes the set of elements in $A$ but not in $B$. Then, the $k$-nearest neighbor ($k$-NN) version of the KL estimator is given by
\begin{equation}\label{k-NN.KL}
H^{k.KL}_{n}=\frac{d}{n}\sum_{i=1}^{n}\log R_{i,k,n-1}+\log \left( \dfrac{\pi^{\frac{d}{2}}}{\Gamma(\frac{d}{2}+1)}\right)-L_{k-1}+\gamma+\log (n-1),
\end{equation}
where, $L_{0}=0$, $L_{j}=\sum_{r=1}^{j}\frac{1}{r}$, and  $\gamma=0.5772\cdots$ denotes Euler's constant.  

  Recently, Al-Labadi et al. (2021) derived a Bayesian nonparametric (BNP) estimator of (\ref{MI-basic}) and used it for the test of independence.  In their paper, Al-Labadi et al. (2021) did not discuss how to numerically implement their estimator. Thus, the main focus of this paper is to assess the numerical estimation of $MI(F)$ through several examples and a comparative study.

The reminder of this paper is as follow. In Section 2, A BNP estimator of mutual information is stated through proposing posterior joint and marginal entropies via Dirichlet process. A computational algorithm is also presented in this section to compute the estimator. In Section 3, the procedure is investigated through several simulation examples and the results are compared to its frequentist counterpart. In Section 4, a real data example is used to show the applicability of the procedure. Finally, the paper is concluded by Section 5.


\section{Bayesina Nonparametric Estimator of Mutual Information}
The BNP estimator of the mutual information (Al-Labadi et. al., 2021) uses the Dirichlet process (Ferguson, 1973) and the  $k$-nearest neighbor estimator of Singh et al. (2003) as described in (\ref{k-NN.KL}).  Let  $DP(a,G)$ be the Dirichlet process with a positive real number $a$ and a fixed probability measure $G$. For some choices of $a$ and $G$, let $F\sim DP(a,G)$. By the conjugacy property of the Dirichlet process, for an observed sample $\mathbf{x}_{d\times n}=(\mathbf{x}_{1},\ldots,\mathbf{x}_{n})$ generated from $F$, the posterior distribution of $F$ given $\mathbf{x}_{d\times n}$, denoted by $F|\mathbf{x}_{d\times n}$, is $DP(a+n,G_{a,n})$, where $G_{a,n}=a(a+n)^{-1}G+n(a+n)^{-1}F_{n}$ and $F_{n}$ is the empirical cumulative distribution function  of the sample
$\mathbf{x}_{d\times n}$. Then, the mutual information of the posterior distribution of $F$ is given by
\begin{align}\label{MIpost}
MI^{pos}&=-H^{pos}_{N,a+n,k}(F)+\sum_{i=1}^{d}H^{pos}_{N,a+n,k}(F_{i}),
\end{align}
where
\begin{align}\label{k-NN.post}
H^{pos}_{N,a+n,k}(F)&=\sum_{i=1}^{N}J_{i,N}\left(\log \frac{(N-1)\pi^{\frac{d}{2}}(R_{,k,N-1}(\mathbf{Y}_{i}))^{d}}{k\Gamma(\frac{d}{2}+1)}\right)-L_{k-1}+\gamma+\log k,
\end{align}
$k\in\lbrace 1,\ldots,N-1\rbrace$, $(J_{1,N},\ldots,J_{N,N})\sim \mbox{Dirichlet}((a+n)/N,$ $\ldots,(a+n)/N)$, $\mathbf{Y}_{1},\ldots,\mathbf{Y}_{N}\overset{i.i.d.}{\sim}G_{a,n}$ and $R_{k,N-1}(\mathbf{Y}_{i})$ is the Euclidean distance between $\mathbf{Y}_{i}$ and its $k$-th closest neighbor. Note that, $H^{pos}_{N,a+n,k}(F_{i})$  can be similarly derived by using $F_{i}|\mathbf{x}_i\sim DP(a+n,G_{i,n,a})$, for $i= 1,\ldots,d$, where $G_{i,n,a}$ is the $i$-th marginal of the cdf  $G_{a,n}$. To improve the estimation Al-Labadi et. al. (2021), we use the \textit{midhinge} of $MI^{pos+}$ as follows. Let $Q_1$ and $Q_3$ be the first and the third quartile of $MI^{pos+}$, respectively. Then, the \textit{midhinge} of $MI^{pos+}$ defined by $MI^{pos+}_{mid}=(Q_1+Q_3)/2$ is used as the BNP estimator for the mutual information. The rationale of using this estimator is described  via a simulation study in Section 3, Table \ref{mean-mid}. Note that, since the closed form of the distribution of $MI^{pos+}$ is not available, the empirical distribution of $MI^{pos+}$ based on samples of size $\ell$ are required to estimate  the posterior midhinge. Clearly, the implementation of the proposed estimation requires considering choices of $a$ and $G$ in $DP(a,G)$. We propose to use $a=0.05$. This small value of $a$ should make the estimated value independent  from any choice of $G$. For simplicity, we set $G=N_{d}(\mathbf{0}_{d},I_{d})$. A detailed  computational algorithm of the proposed estimator is  presented below.

\noindent\textbf{Algorithm 1:} \textit{BNP estimation of mutual information}
\begin{small}
\begin{enumerate}
\item Set $a$ as small as possible, say $a=0.05$.
\item Let $G$ be the cdf of $N(\mathbf{0}_{d},I_{d})$ and  generate a sample from $DP(a+n,G_{a,n})$ as follows:
\begin{enumerate}
\item[i.] Fix a large positive integer $N$ and generate i.i.d. $\mathbf{Y}_{i}\sim G$, for $i=1,\ldots,N$.
\item[ii.] To generate $(J_{i,N})_{1\leq i\leq N}$, put $J_{i,N}=\Gamma_{i,N}/\sum_{i=1}^{N}\Gamma_{i,N}$, where $(\Gamma_{i,N})_{1\leq i\leq N}$ is a sequence of i.i.d. $Gamma((a+n)/N,1)$ random variables independent of $\mathbf{Y}_{i}$.
\item[iii.] Return $P_{N}=\sum_{i=1}^{N}J_{i,N}\delta_{\mathbf{Y}_{i}}$.
\end{enumerate}
\item For the   sample generated in the previous step, use \eqref{k-NN.post} to compute $H^{pos}_{N,a+n,k}(F)$ and $H^{pos}_{N,a+n,k}(F_{i})$, for $i=1,\ldots,d$.
\item
Use $H^{pos}_{N,a+n,k}(F)$ and $H^{pos}_{N,a+n,k}(F_{i})$'s in \eqref{MIpost} to compute $MI^{pos}$, and then $MI^{pos+}$.
\item
Repeat steps (1)-(4) to obtain a sample of $\ell$ values from $MI^{pos+}$.
\item
Compute the 0.25-th and 0.75-th quantile of $\ell$ values generated in step (5), denoted by $Q_1$ and $Q_3$, respectively. Deliver $\dfrac{Q_1+Q_3}{2}$ as the estimator of mutual information.
\end{enumerate}
\end{small}
\section{Simulation studies}
The performance of the BNP methodology in estimating mutual information is evaluated through several illustrative examples including $d$-variate distributions; normal, $t$-student with $df$ degrees of freedom, and Maxwell-Boltzmann distributions. We generate $r = 1000$ samples from each distribution with $n = 20, 30, 50$. Next, we compute $MI_{mid}^{pos+}$ and $(MI_{mid}^{pos+}-MI^{T})^{2}$  for each of the generated sample, where $MI^{T}$  denotes the true value of mutual information. Over the $r$ samples, we record the average value of  $MI_{mid}^{pos+}$  and the average value of $(MI_{mid}^{pos+}-MI^{T})^{2}$, where the later average is the mean squared error (MSE). For the sake of comparison, over the $r$ samples, the average of mutual information estimation of Berrett \& Samworth (2019) based on the weighted version of the KL estimator ($MI^{W.KL}$) and the MSE of $MI^{W.KL}$  are reported. The $\mathsf{R}$ package \textbf{IndepTest} is used to compute $MI^{W.KL}$.  Hereafter, let $c_{\mathbf{d}}$ be the $d$-dimensional column vectors of $c$'s, $I_d$ be the $d\times d$ identity matrix, $A_d=(a_{ij})_{1\leq i,j\leq d}$ be the $d\times d$ matrix with 1's on the main diagonal, $a_{d,d-1}=a_{d-1,d}=0.5$ and $0$'s elsewhere, $B_d$ be the $d\times d$ matrix with $1$'s on the main diagonal and $0.9$'s elsewhere,
$\Sigma_d$ be the $d\times d$ matrix with $(1,2,1,\ldots,1)$ on the main diagonal and $0.5$'s elsewhere.
The following notations have been used:
$N_{d}(\mathbf{0}_{d},\Sigma_{d})$ for a $d$-variate normal distribution with mean vector $\mathbf{0}_{d}$ and covariance matrix $\Sigma_{d}$, and $MI^T=\frac{d}{2}\sum_{i=1}^{d}(\log(2\pi e\sigma^{2}_{i}))-\frac{1}{2}\log((2\pi e)^{d}\det(\Sigma))$, where $\sigma^{2}_{i}$ is the $i$-th diagonal element of $\Sigma_d$,
$t_{df}(\mathbf{0}_{d},I_{d})$ for a $d$-variate $t$-student distribution with location parameter $\mathbf{0}_{d}$, scale parameter $I_{d}$ and $df$ degrees of freedom, and $MI^T=d\big(\frac{df+1}{2}[\psi((1+df)/2)-\psi(df/2)]+\log[\sqrt{df}B(df/2,1/2)]\big)-\big\lbrace-\log\frac{\Gamma((df+d)/2)}{\Gamma(df/2)(df\pi)^{d/2}}+\frac{df+d}{2}[\psi(\frac{df+d}{2})-\psi(\frac{df}{2})]\big\rbrace$, where $B(\cdot,\cdot)$ denotes the  beta function,
$SP_d(LN(0,0.25))$ for a $d$-variate spherical distribution with lognormal distribution $LN(0,0.25)$ for radii,
$F_{1}\otimes\ldots\otimes F_{d}$ for a $d$-variate distribution with $d$ independent marginal distributions $F_{1},\ldots,F_{d}$,
and
$Mwell(\mathbf{c}_{d})=Mwell(c)\otimes\cdots\otimes Mwell(c)$, where $Mwell(c)$  denotes the Maxwell-Boltzman distribution with scale parameter $c$ and $MI^T=0$.

\begin{figure}[ht]
 \centering
    \subfloat[$N_4(\mathbf{0}_{4},\Sigma_4)$, $MI^T=0.450$]{{\includegraphics[width=6cm]{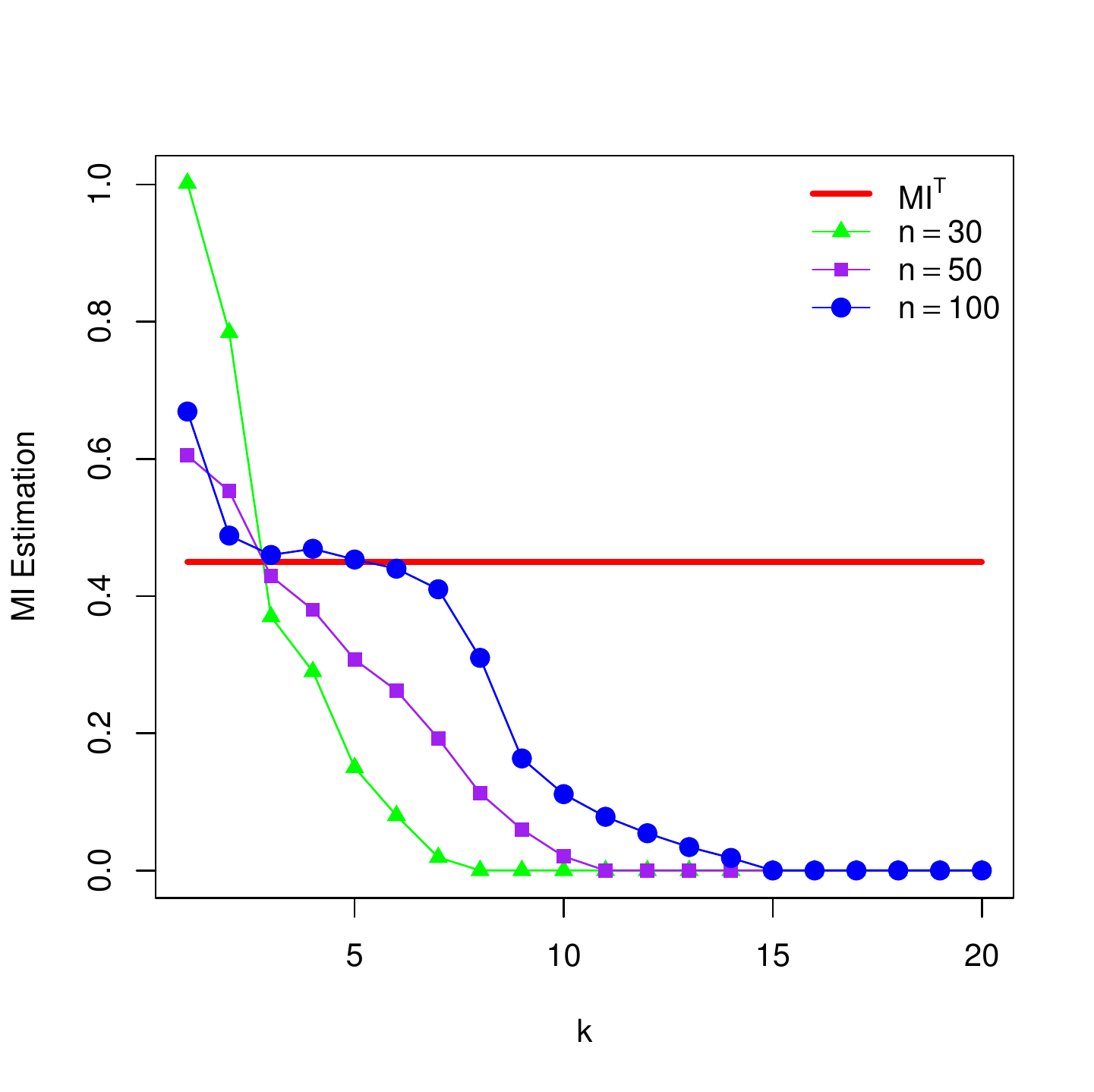} }}%
    \qquad
    \subfloat[$t_{3}(\mathbf{0}_{4},I_{4})$, $MI^{T}=0.195$]{{\includegraphics[width=6cm]{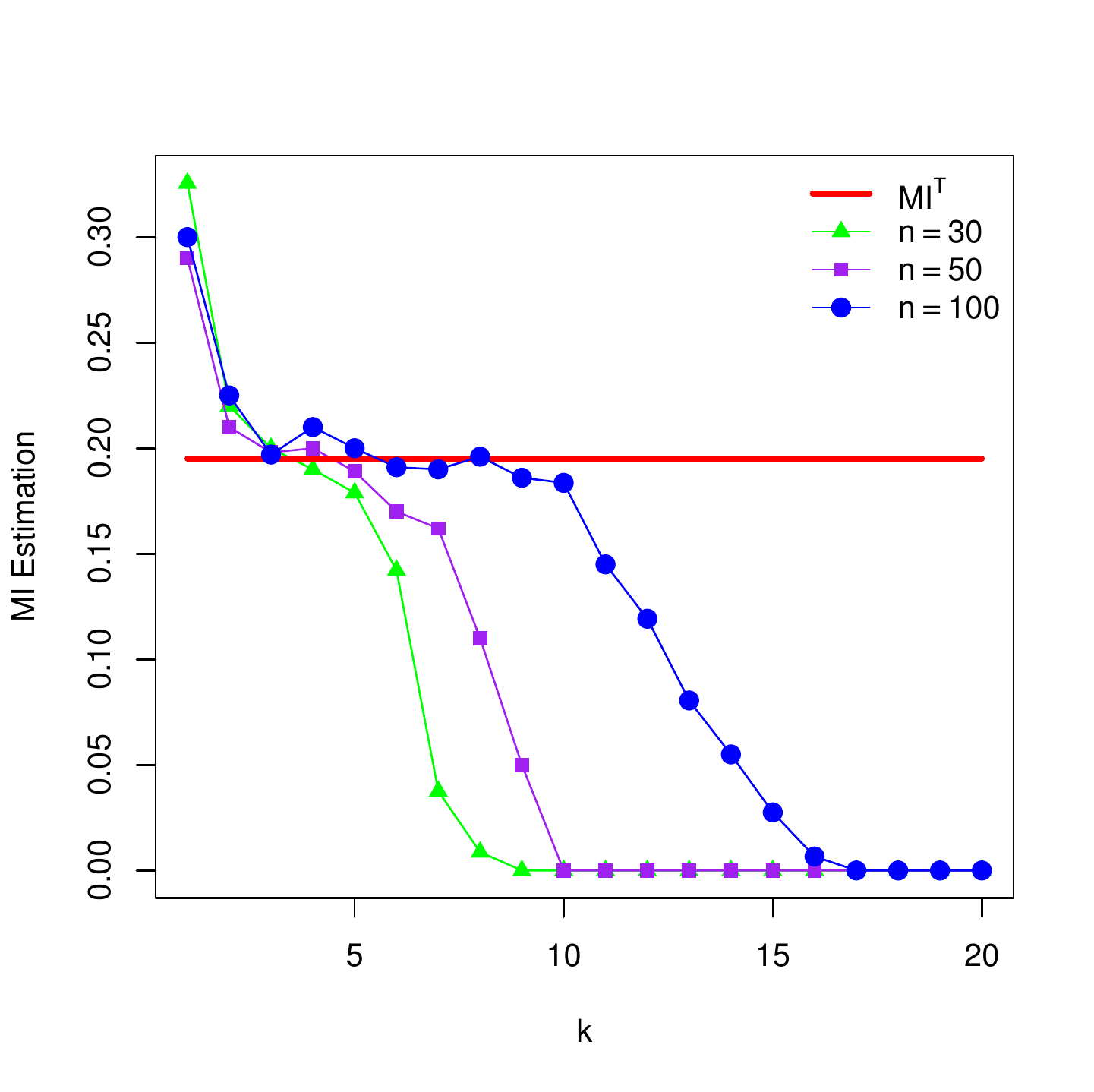} }}
    \caption{ The average values of the BNP mutual information estimation over $r$ samples with $k=1,\ldots,20$ for various sample sizes ($n=30,50,100$).}\label{sen-k}%
\end{figure}

\begin{figure}[ht]
 \centering
    \subfloat[Plot of the MI estimation]{{\includegraphics[width=6cm]{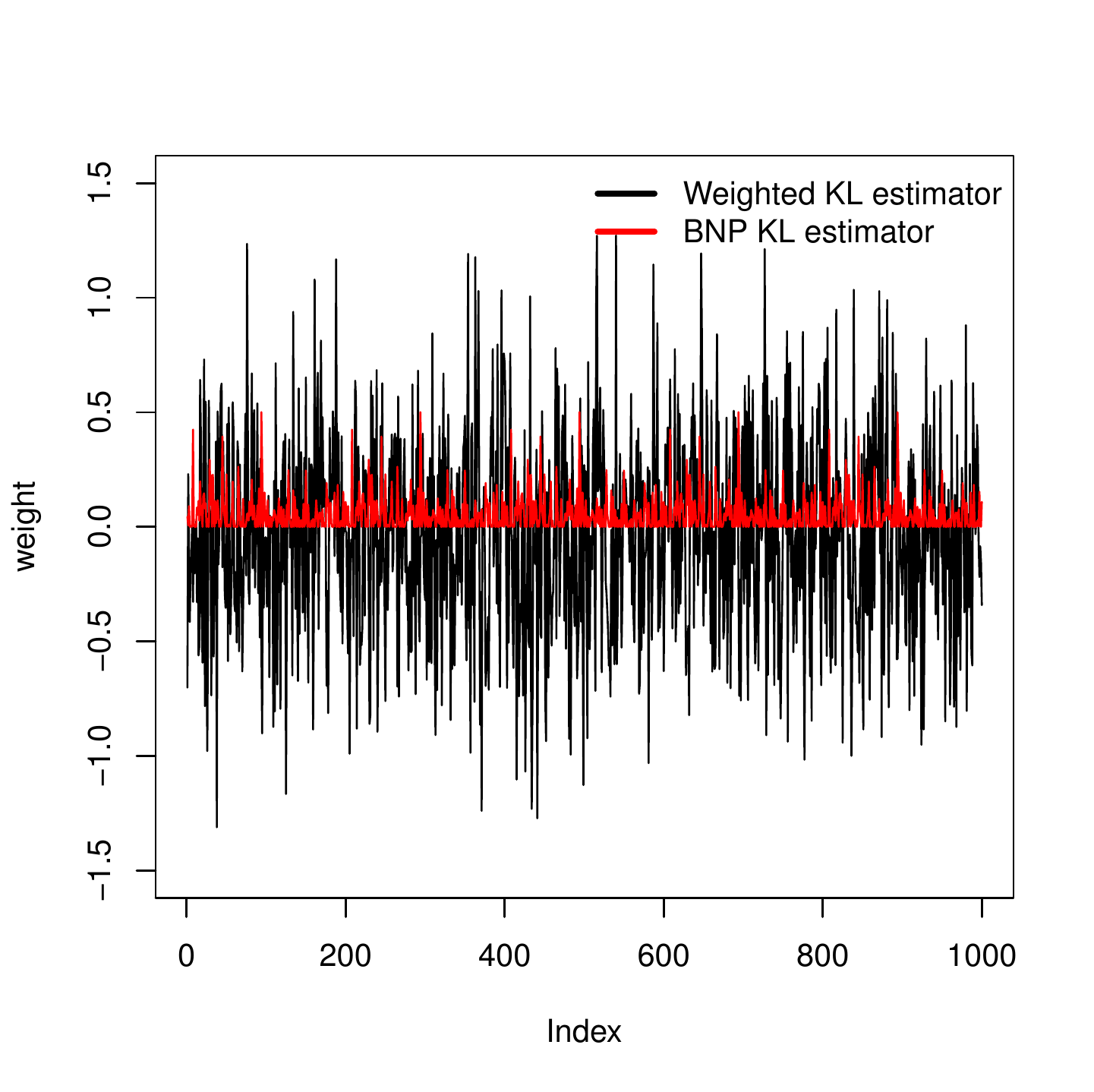} }}%
    \qquad
    \subfloat[Density of the MI estimation]{{\includegraphics[width=6cm]{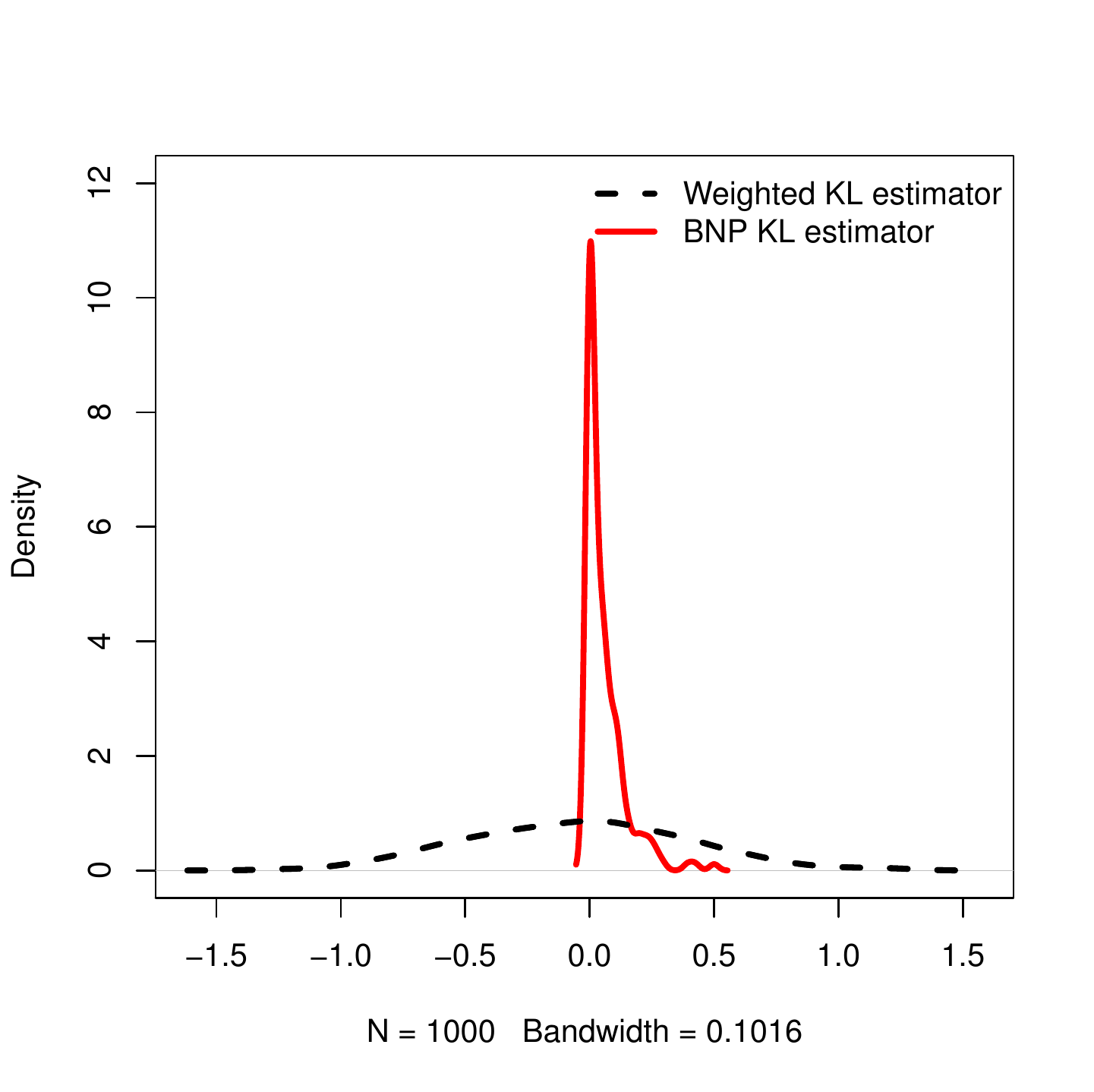} }}
    \caption{The BNP estimations of mutual information for $r$ samples generated from $N_4(\mathbf{0}_4,I_4)$ with $n=50$ and $k=3$.}%
    \label{common.dist}%
\end{figure}

To continue, we first compare $MI^{pos}$  and $MI^{pos+}$ as potential estimators  of \eqref{MI-basic}. For this, $r$ samples of size $50$ are generated from $N_4(\mathbf{0}_4,I_4)$ and $N_4(\mathbf{0}_4,\Sigma_4)$, and then the average of the mean of $\ell=1000$ values of $MI^{pos}$ and $MI^{pos+}$ (posterior mean)  over $r$ samples are presented in Table \ref{mean-mid} against the average value of the midhinge of $\ell$ values of $MI^{pos}$ and $MI^{pos+}$ (posterior midhinge) over $r$ samples. It follows from Table \ref{mean-mid} that the posterior midhinge of $MI^{pos+}$ is a range-preserving Bayesian estimator and is not affected by outliers (compare the gray column in Table \ref{mean-mid} to other columns).
\begin{table}[ht]
\center
\setlength{\aboverulesep}{0pt}
\setlength{\belowrulesep}{0pt}
\setlength{\extrarowheight}{1.1 mm}
\setlength{\tabcolsep}{2 mm}
\caption{The average values of the posterior mean and midhinge of $MI^{pos}$ and $MI^{pos+}$ over $r$ samples from $N_4(\mathbf{0}_4,I_4)$ and $N_4(\mathbf{0}_4,\Sigma_4)$ and their relevant MSEs with $k=3$.}
\scalebox{.8}
{
\begin{tabular}{cccccc}
\toprule
\multirow{2}[3]{*}{\bfseries Examples}&\multirow{2}[3]{*}{$MI^{T}$}& \multicolumn{2}{c}{$MI^{pos}$}&\multicolumn{2}{c}{$MI^{pos+}$} \\\cline{3-4}\cline{5-6}
&&mean(MSE)&midhinge(MSE)&mean(MSE)&midhinge(MSE)\\
\hline
$N_4(\mathbf{0}_4,I_4)$&$0$&$-0.097(0.0479)$&$-0.089(0.0465)$&$0.090(0.0198)$&\cellcolor{gray!10}$0.053(0.0105)$\\
$N_4(\mathbf{0}_4,\Sigma_4)$&$0.45$&$0.327(0.0723)$&$0.338(0.071)$&$.0.384(0.0389)$&\cellcolor{gray!10}$0.401(0.0362)$\\
\bottomrule
\end{tabular}
}\label{mean-mid}
\end{table}
It is also interesting to check the effect of the choice of $k$ on the posterior midhinge of $MI^{pos+}$. Figure \ref{sen-k} shows the average values of $MI_{mid}^{pos+}$ over $r$ samples generated from $N_4(\mathbf{0}_{4},\Sigma_{4})$ and $t_{3}(\mathbf{0}_{4},I_{4})$ with $k=1,\ldots,20$ for various sample sizes ($n=30,50,100$). In all cases, it seems that $k=3$ is a suitable choice. However, increasing the value of $k$ increases the error of estimation. This fact follows the theoretical result presented by Al-Labadi et al. (2021, Cor. 2).
The sensitivity of the BNP estimation of mutual information to the choice of $a$ and $G$ is also  investigated by Table \ref{sen-a.G}. We reported the BNP estimation for samples generated from two distributions $N_3(\mathbf{0}_{3},A_{3})$ and $t_{3}(\mathbf{0}_{3},I_{3})$  with $n=30$ based on various choices of $a$ and $G$. It is obvious that using different $G$ with $a=0.05$ has no considerable impact on estimated values (see the gray column in Table \ref{sen-a.G}). That is, the estimated values do not depend on the choice of $G$ when $a=0.05$. On the other side, large values of  $a$  (such as $a=5$) leads to increase the effect of the choice of $G$ on estimated values.
\begin{table}[H]
\center
\setlength{\aboverulesep}{0pt}
\setlength{\belowrulesep}{0pt}
{\setstretch{.6}
\setlength{\tabcolsep}{2 mm}
\caption{The average values of the posterior mean and midhinge of $MI^{pos}$ and $MI^{pos+}$ over $r$ samples from $N_4(\mathbf{0}_4,I_4)$ and $N_4(\mathbf{0}_4,\Sigma_4)$ and their relevant MSEs with $k=3$.}\label{sen-a.G}
\scalebox{.8}
{
\begin{tabular}{cccccc}
\toprule
\multirow{2}[3]{*}{\bfseries Example}&\multirow{2}[3]{*}{$MI^{T}$}&\multirow{2}[3]{*}{$G$}& \multicolumn{3}{c}{$MI_{mid}^{pos+}(MSE)$}\\\cline{4-6}
&&&$a=0.05$&$a=5$&$a=10$\\
\hline
$N_3(\mathbf{0}_{3},A_3)$&0.143&$N_{3}(\mathbf{0}_{3},I_{3})$&\cellcolor{gray!10}$0.141(0.0193)$&$0.11(0.0296)$&$0.09(0.0311)$\\
&&$N_{3}(\mathbf{3}_{3},B_{3})$&\cellcolor{gray!10}$0.148(0.0188)$&$1.211(1.3785)$&$1.801(2.5853)$\\
&&$SP_{3}(LN(0,0.25))$&\cellcolor{gray!10}$0.140(0.0186)$&$0.321(0.0412)$&$0.394(0.0452)$\\
\hline
$t_3(\mathbf{0}_{3},I_3)$&0.110&$N_{3}(\mathbf{0}_{3},I_{3})$&\cellcolor{gray!10}$0.161(0.0237)$&$0.103(0.0345)$&$0.081(0.0389)$\\
&&$N_{3}(\mathbf{3}_{3},B_{3})$&\cellcolor{gray!10}$0.159(0.0221)$&$0.988(1.2531)$&$1.412(1.9356)$\\
&&$SP_{3}(LN(0,0.25))$&\cellcolor{gray!10}$0.163(0.0240)$&$0.410(0.0821)$&$0.423(1.0911)$\\
\bottomrule
\end{tabular}
}}
\end{table}
Now, through different scenarios, the performance of $MI_{mid}^{pos+}$ to estimate the mutual information is examined in Table \ref{comm.dis.T}. By comparing the column of $MI_{mid}^{pos+}$ with the column of $MI^{W.KL}$, it is seen that $MI_{mid}^{pos+}$ has a smaller MSE. For instance, in Table \ref{comm.dis.T}, when $N_4(\mathbf{0}_4,I_4)$ and $n=50$ , the average values of $MI_{mid}^{pos+}$ over $r$ samples is $0.053$, while the average values of $MI^{W.KL}$ is $-0.033$. This shows that $MI^{W.KL}$ is not a range-preserving estimator of mutual information. Also, the corresponding MSE of $MI_{mid}^{pos+}$  is $0.0105$ which is  smaller than the MSE of $MI^{W.KL}$ ($0.2069$). Additional,  for this example,  the plot of the density of $MI_{mid}^{pos+}$  against that of $MI^{W.KL}$  is presented  in Figure \ref{common.dist}. It follows from this figure  that the  concentration of the BNP estimation around $MI^{T}$  is better than its frequentist counterpart.
\begin{table}[H]
\center
\setlength{\aboverulesep}{0pt}
\setlength{\belowrulesep}{0pt}
\setlength{\extrarowheight}{1.5 mm}
{\setstretch{.6}
\setlength{\tabcolsep}{5 mm}
\caption{The average values of the BNP mutual information estimation and $RB(Str)$ over $r$ samples and its relevant MSE under several distributions with $a=0.05$ and $k=3$.}\label{comm.dis.T}
\scalebox{.8}
{
\begin{tabular}{lcccccc}
\toprule
\multirow{2}[3]{*}{\bfseries Example}&$d$&\multirow{2}[3]{*}{$n$}&\multicolumn{2}{c}{BNP}&\multicolumn{2}{c}{Berrett \& Samworth}\\\cmidrule(lr){4-5}\cmidrule(lr){6-7}
&$(MI^{T})$&&$MI_{mid}^{pos+}$&$MSE$&$MI^{W.KL}$&$MSE$\\\hline
$N_{d}(\mathbf{0}_{d},I_{d})$&2&20&$0.084$&$0.0207$&$-0.063$&$0.0626$\\
&(0)&30&0.075&$0.0139$&$-0.011$&$0.0385$\\
&&50&$0.059$&$0.0070$&$0.008$&$0.0270$\\\cmidrule(lr){2-7}
&4&20&$0.047$&$0.0117$&$-0.149$&$0.5049$\\
&(0)&30&$0.050$&$0.0088$&$-0.103$&$0.3293$\\
&&50&$0.053$&$0.0105$&$-0.033$&$0.2069$\\\cmidrule(lr){1-7}
$N_{d}(\mathbf{0}_{d},\Sigma_{d})$&2&20&$0.109$&$0.0208$&$0.008$&$0.0627$\\
&(0.066)&30&$0.108$&$0.0146$&$0.032$&$0.0452$\\
&&50&$0.106$&$0.0082$&$0.072$&$0.0305$\\\cmidrule(lr){2-7}
&4&20&$0.294$&$0.0591$&$0.294$&$0.5431$\\
&(0.450)&30&$0.337$&$0.0542$&$0.376$&$0.3890$\\
&&50&$0.401$&$0.0362$&$0.397$&$0.2380$\\\cmidrule(lr){1-7}
$N_{d}(\mathbf{0}_{d},A_{d})$&2&20&$0.177$&$0.0295$&$0.079$&$0.0729$\\
&$(0.143)$&30&$0.177$&$0.0198$&$0.110$&$0.0505$\\
&&50&$0.160$&$0.0116$&$0.170$&$0.0314$\\\cmidrule(lr){2-7}
&4&20&$0.071$&$0.0192$&$-0.002$&$0.4842$\\
&$(0.143)$&30&$0.090$&$0.0180$&$0.055$&$0.3631$\\
&&50&$0.129$&$0.0124$&$0.091$&$0.2048$\\\cmidrule(lr){1-7}
$t_{3}(\mathbf{0}_{d},I_{d})$&2&20&$0.095$&$0.0269$&$0.016$&$0.0593$\\
&$(0.042)$&30&$0.087$&$0.0259$&$0.077$&$0.0426$\\
&&50&$0.083$&$0.0147$&$0.086$&$0.0290$\\\cmidrule(lr){2-7}
&4&20&$0.161$&$0.0457$&$-0.097$&$0.5368$\\
&(0.195)&30&$0.218$&$0.0425$&$-0.023$&$0.3236$\\
&&50&$0.211$&$0.0378$&$0.057$&$0.2243$\\\cmidrule(lr){1-7}
$t_{20}(\mathbf{0}_{d},I_{d})$&2&20&$0.080$&$0.0171$&$-0.053$&$0.0673$\\
&(0.001)&30&$0.087$&$0.0186$&$-0.019$&$0.0411$\\
&&50&$0.075$&$0.0114$&$0.015$&$0.0284$\\\cmidrule(lr){2-7}
&4&20&$0.061$&$0.0151$&$-0.144$&$0.5031$\\
&(0.006)&30&$0.082$&$0.0192$&$-0.100$&$0.3379$\\
&&50&$0.081$&$0.0161$&$-0.045$&$0.1942$\\\cmidrule(lr){1-7}
$Mwell(\mathbf{10}_{d})$&2&20&$0.080$&$0.0210$&$-0.053$&$0.0593$\\
&(0)&30&$0.065$&$0.0127$&$-0.033$&$0.0412$\\
&&50&$0.060$&$0.0141$&$-0.014$&$0.0304$\\\cmidrule(lr){2-7}
&4&20&$0.045$&$0.0109$&$-0.132$&$0.5151$\\
&(0)&30&$0.049$&$0.0127$&$-0.099$&$0.3464$\\
&&50&$0.055$&$0.0094$&$-0.056$&$0.2029$\\\cmidrule(lr){1-7}
\bottomrule
\end{tabular}
}}
\end{table}
\section{Application}
The combined cycle power plant (CCPP) data set is considered to evaluate the applicability of the BNP method on a real data set. This data set contains 9568 five-dimensional data points. It is collected from 2006 to 2011 and is available at \url{https://archive.ics.uci.edu/ml/datasets/combined+cycle+power+plant}. Here, the goal is to predict the net hourly electrical energy output of the plant based on the temperature (T), the ambient pressure (AP), the relative humidity (RH) and the exhaust vacuum (V). Thus, it is significant to quantify the amount of dependence between the four variables T, AP, RH, and V are independent. In addition, besides using all 9568 data points, we considered three samples with sample sizes $n=20,30$ and $50$ generated randomly from the whole data set. The proposed method then is implemented. The values of the BNP mutual information estimation are presented in Table \ref{real.T}, which indicates a certain value of the mutual dependence between variables.
\begin{table}[h]
\center
\setlength{\aboverulesep}{0pt}
\setlength{\belowrulesep}{0pt}
\setlength{\extrarowheight}{1.5 mm}
\setlength{\tabcolsep}{5 mm}
\caption{The values of the BNP mutual information estimation for CCPP data set with $a=0.05$, $k=3$ and various sample sizes $n$.}\label{real.T}
\scalebox{.8}
{
\begin{tabular}{lcccc}
\toprule
&$n=20$&$n=30$&$n=50$&$n=9568$\\
\cmidrule(lr){2-5}
$MI_{mid}^{pos+}$&0.457&0.481&0.482&0.779\\
\bottomrule
\end{tabular}
}
\end{table}
\section{Conclusion}
A Bayesian nonparametric estimation of the mutual information has been presented by using of Dirichlet process and $k$-nearest neighbor distance. The procedure has been investigating by several simulation study examples where the results reflect  the good performance of the procedure to estimate the mutual information with a small MSE. Finally, a real data example has been investigated to reveal the applicability of the procedure.

\end{document}